\documentclass[12pt,preprint]{aastex}
\renewcommand{\baselinestretch}{1.}
\usepackage{graphicx}
\usepackage{fullpage,amsmath,amssymb,latexsym}
\usepackage{multirow}
\usepackage{natbib}
\usepackage{url}
\usepackage{graphicx}

\shorttitle{The fuzziness of Jupiter's core}
\shortauthors{Helled \& Stevenson}

\begin{document}

\title{The fuzziness of giant planets' cores}
\author{Ravit Helled$^{1,2}$ \& David Stevenson$^3$}
\affil{
$^1$Institute for Computational Science, 
University of Zurich, Switzerland.\\ 
$^2$School of Geosciences, 
Tel-Aviv University, Israel.\\
$^3$Division of Geological
and Planetary Sciences, Caltech, USA.}

\begin{abstract}
Giant planets are thought to have cores in their deep interiors, and the division into a heavy-element core and hydrogen-helium envelope is applied in both formation and structure models. We show that the primordial internal structure depends on the planetary growth rate, in particular, the ratio of heavy elements accretion to gas accretion. For a wide range of likely conditions, this ratio is in one-to-one correspondence with the resulting post-accretion profile of heavy elements within the planet. This flux ratio depends sensitively on the assumed solid surface density in the surrounding nebula. We suggest that giant planets' cores might not be distinct from the envelope and includes some hydrogen and helium, and the deep interior can have a gradual heavy-element structure.  Accordingly,  Jupiter's core may not be well-defined. 
Accurate measurements of Jupiter's gravitational field by {\it Juno} could put constraints on Jupiter's core mass. 
However, as we suggest here, the definition of Jupiter's core is complex, and the core's physical properties (mass, density) depend on the actual definition of the core and on its growth history. 
\end{abstract}
%\par

%\clearpage
\section{Introduction}
%In the standard model of giant planet formation, core accretion, first a solid core is formed until accretion of gaseous becomes efficient enough to follow rapid gas accretion. 
The formation of a giant planet in the core accretion model can be divided into three main phases \citep{Pollack1996,Helled2014b}:
\begin{itemize}
	\item {\bf phase-1}: Primary core/heavy-element accretion. During this phase, the core accretes solids within its feeding zone, until it reaches isolation mass. The envelope's mass (hydrogen-helium) is much smaller than the heavy-element mass $M_Z$. Solid material could arrive as small bodies (pebbles) or as very large bodies (merging embryos/giant impacts) but here we focus on the intermediate-case which is also the best-studied case - planetesimal accretion. During this early phase, the protoplanet's atmosphere is close to hydrostatic equilibrium and merges smoothly with the low-density nebula at the Hill Sphere, and most of the accreted planetesimals reach surface of a well-defined heavy element core. 
	\item {\bf phase-2}: Slow envelope/gas accretion. During this phase, the solid accretion rate decreases, and the gas (hydrogen and helium) accretion rate increases until the envelope accretion rate exceeds the core (solids) accretion rate. The envelope's growth expands the planet's feeding zone and thus allows more planetesimals to be accreted but at a slow rate. 
  \item {\bf phase-3}: Rapid gas accretion. Once the protoplanet reaches crossover mass ($M_{H-He}=M_{Z}$), gas accretion rate continuously increases, and exceeds the solid accretion rate until the disk can no longer supply gas fast enough to maintain equilibrium and keep up with the planetary contraction. After reaching that point, hydrodynamic gas accretion begins. 
\end{itemize}
\par

The disruption of planetesimals breaks them to small particles. Above $\sim$1 M$_{\oplus}$, the accretional energy (per unit-mass) exceeds the latent heat of vaporization of rock. %(indeed, EarthÕs energy of formation is about equal to the energy to vaporize Earth entirely). 
The energy required to keep the gaseous atmosphere hot is small by comparison. At the earliest stages ($\sim$a few M$_{\oplus}$ or less), the gaseous mass in the atmosphere is small and increases with time, in order to maintain hydrostatic equilibrium. As a result, the heavy element mass influx greatly exceeds the gas mass influx and the rock and ice still form a ÒcoreÓ, albeit very hot, even supercritical (with a small amount of gas mixed in). 
At a later stage (phase-2), the main focus here, the ratio of gas mass influx to heavy element influx rises, eventually reaching and exceeding unity. Most of the mass ($\sim$90\% of the planet) accumulates during phase-3 (runaway) when the gas influx is fast and thus greatly exceeds the heavy element influx.
\par

The planetary primordial internal structure depends on various physical processes and model assumptions. While various detailed numerical models of giant planet formation exist, there is an advantage in using simple models in order to get a feeling for the problem. Therefore, our arguments build up on the model of \citet{Stevenson1982} where the surface temperature of a  growing core is $\sim$4000K at one Earth mass (M$_{\oplus}$) and is increasing as $\propto M^{2/3}$. 
The gas' density at that surface depends on the accretion rate and opacity, for an accretion timescale of 10$^6$ years, it's $\sim$10$^{-3}$ g/cm$^3$, suggesting that the atmospheric mass is very small but the ram pressure exerted on incoming planetesimals of velocity $v$ is $\sim$1000bar.(v/10km/s)$^2$, sufficient for planetesimal disruption. % the planetesimal. 
\par 

We claim that the post-accretion structure of the planetary core accretion directly reflects the history of relative accretion (gas accretion rate to heavy-element accretion rate):
%That is, 
\begin{equation}
Z(m) \approx (dM_Z/dt)/(dM_{tot}/dt) \quad \text{at} \quad M(t)=m,
\end{equation}
%Z(m) Å (dMZ/dt)/(dMtot/dt) at M(t)=m
where $Z(m)$ is the post-accretion ratio of heavy elements to gas (hydrogen and helium) at the radius that contains mass $m$. $M_Z$ is the total mass of heavy elements, and $M_{tot}$ is the total mass, both evaluated at the time when the total mass is m$\ll$ final Jupiter mass. This means the heavy element distribution post-accretion (at an age of several million years) is a direct reflection of how the planet accreted. This isn't necessarily the present day structure because convective mixing can spread the heavy elements further \citep{Leconte2012,Vazan2016}. Accordingly, {\it the minimum fuzziness of the core of a giant planet is a direct consequence of the gradual change in the accretion ratio of heavy elements and hydrogen and helium over time}. This would only go away if one thought (unrealistically) that the planet abruptly switched from pure heavy elements to pure gas accretion at some time. This is impossible since quasi-hydorstatic equilibrium mandates gas accretion when heavy elements are accreted.
\par 

Our suggestion is linked to the following criteria:
\begin{enumerate} 
\item A monotonic decrease of $(dM_Z/dt)/(dM_{tot}/dt)$ with increasing $m$. 
\item Heating due to the accretion causes a decrease in local gas density that is less than the increase of density arising from the injection of heavy elements into the same mass of gas.
\item Deposition of incoming heavies occurs over a range of radii that corresponds to a small mass fraction of the planet at that time.
\item Deposition occurs at a radius that encloses most of the mass at that time. 
\end{enumerate} 

The first criterion is required for gravitational stability of the resulting $Z$-profile. It's satisfied by all standard models of planetary growth.  
The second depends on the dimensionless parameter $A$ defined by the ratio of density change due to heating to the density change due to increase in the mean molecular weight:
\begin{equation}
	A=v^2/2C_pT
\end{equation}
where $v$ is the velocity at the location where planetesimal break-up occurs, $T$ the ambient temperature at that radius, and $C_p$ the specific heat. This assumes the heavies have a mean  molecular weight far larger than that of a hydrogen-helium mixture. 
A necessary condition for vigorous mixing is $A$ substantially greater than unity. In that case , the consequence of an incoming planetesimal would be a rising plume of very hot but enriched gas 
 %(cf. the Shoemaker-Levy impacts)
  rather than the "local" deposition of new heavy material. 
Even at $A \sim 1$, mixing isn't highly efficient. Actual models predict $A < 1$ though not by a large amount. Typically, $v < $10 km/s (break-up occurs at $\sim$10$^{10}$ cm when the total mass is $\sim$10 M$_{\oplus}$). 
\par

Criteria 3 and 4 are only needed to sharpen the validity of Equation 1. They're typically satisfied but the essence of our main claim would still apply even if they weren't strictly satisfied. 
However, validity of our suggestion regarding $Z(m)$ doesn't automatically lead to a prediction for the final $Z$-profile which depends on the specific conditions under which the giant planet has formed, in particular, the solid-surface density $\sigma$.

\section{Jupiter's Primordial Internal Structure}
In order to demonstrate the sensitivity of the internal structure to the formation history,  we use two formation models for Jupiter: Model-1 with $\sigma=10$ g cm$^{-2}$ and Model-2 with $\sigma=6$ g cm$^{-2}$ at 5.2 AU with planetesimal sizes of 100 km taken from \citet{Lozovsky2017}.
\par

The calculation begins with a small core ($\sim$ 0.1 M$_{\oplus}$). The heavy-element distribution is determined by following the planetesimals' trajectory as they pass through the envelope accounting for gas drag and gravitational forces. The effects of heating, ablation and planetesimal fragmentation and settling due to saturation are also included. 
In Model-1 crossover is reached after 0.94 Myr, when $M_Z$=16 M$_{\oplus}$, while for Model-2, crossover time is 1.54 Myr and $M_Z$=7.5 M$_{\oplus}$.  % In these models, t
Crossover time is calculated assuming that all the accreted planetesimals reach the core, while the deposition of heavies into the envelope is expected to decrease the time by a factor of a few \citep{Venturini2016}. 
\par

Figure 1 shows $Z(m)$ versus time for the two models up to crossover time.  Jupiter's formation in Model-1 is faster due to the higher $\sigma$, and $M_Z$ is higher than in the case of $\sigma=6$ g cm$^{-2}$ where the primordial structure is more gradual in composition. 
Once the heavy elements are deposited in the atmosphere, the temperature increases significantly due to the change in opacity, and heavy elements evaporate in the envelope. In Model-2 the innermost region of the planet doesn't have a core+envelope structure as in Model-1, but the concentration of heavy elements is very high, mimicking a diluted core. 
\par

Figure 2 shows $Z$ versus total planetary mass throughout the planetary growth until Jupiter's mass is reached. 
%\par
It's interesting to note the large difference in the predicted composition for planets with $\sim$ 20 M$_{\oplus}$ for the two models - while one is composed of mostly heavy elements, and is more similar to Uranus/Neptune, the low-sigma case results in a planet with a much lower mean density, as several of the observed exoplanets. 
In Model-1, the protoplanet consists of mostly heavy elements up to $M\sim$ 11 M$_{\oplus}$, then, as the planet's mass increases, more H-He is accreted and Z drops. Once runway gas accretion begins, $Z$ decreases even further, and the planet's composition is %essentially 
determined by the composition of the accreted gas and the solid accretion rate during phase-3.  
\par

In Model-2, $M_Z$ is smaller and the growth is slower. Therefore, the protoplanet consists of pure-Z up to a smaller mass of $\sim$ 5 M$_{\oplus}$ and $Z$ drops faster than in Model-1. 
In both cases, when the planet reaches a Jupiter-mass $Z$ is nearly solar. 
Model-1 represents a standard core+envelope configuration, while Model-2 results in a less-conventional picture of Jupiter's interior with a more gradual internal structure, and its innermost regions can be viewed as an extended core. The latter corresponds to an onion-like internal structure as suggested by \citet{Stevenson1982}. If Jupiter's interior is found to be more similar to that of Model-2, it would suggest that the local surface density was comparable to the one predicted by the minimum mass solar nebula (MMSN).  During phase-2, the accreted material is mostly H-He gas, but also a small portion of heavy-elements resulting in a slight increase of M$_Z$ when the planet reaches runaway (phase-3). 
%Figure 2 shows the rate of heavy-element mass accretion vs.~time for the two models ({\bf do we want to show it?}). 
\par

Figure 3 shows $Z$ versus normalized mass for the two models at time of 0.66 Myr.
%The results are shown for a model with $\sigma_s$ = 10 g cm$^{-2}$. 
At this time, in Model-1 the total mass is almost 18 M$_{\oplus}$, with 13 M$_{\oplus}$ of heavies while for Model-2 M$_Z$ = 5.6 M$_{\oplus}$ out of a total mass of 6.3 M$_{\oplus}$. 
Since the growing planet in the two models has different formation path, there is no point in comparing the $M_Z/M_{H-He}$, but instead, the actual distribution of the heavy elements. 
It's clear from the figure that Model-1 has a core+envelope structure while in Model-2, the heavy elements are distributed gradually and there is no discontinuity in the density profile. 

If one defines the core of the planet by region with say,  $Z > $0.7, the core is more extended (up to 20\% of the total mass), and also consists of non-negligible amounts of hydrogen and helium. This leads to a lower core density than in the core+envelope configuration. Also, the gradual heavy-element distribution would inhibit large-scale convection. This can affect the heat transport efficiency and therefore also its long-term evolution and current-state internal structure. 
\par

The presented heavy-element distribution assumes that the composition gradients persists during the formation process even in convective regions. Note that this is different from the model of Lozovsky et al. (2007), where the regions with composition gradients that were found to be convective according to the Ledoux convection criteria were {\it assumed} to homogenize instantaneously due to mixing. However, homogenizing convective regions with composition gradients requires a fairly long time $\sim 10^9$ yrs $\gg$ formation timescale (see Vazan et al., 2016). The exact timescale depends on the mixing model, particularly, on the mixing length when using mixing length theory. This requires a detailed study on mixing in giant planets which is beyond the scope of this letter, and we hope to address this topic in detail in future research.   
\par

For the arguments presented here, the core could still be quite well-defined as the central concentration of heavy elements, provided it is central enough. For example, suppose an accretion model (perhaps subsequently modified by convection) predicts that 
$Z(m) =(2a/ \sqrt{\pi})(1-Z_e)exp(-a^2m^2/m_c^2)+Z_e$ where $Z_e$ is $Z$ in the envelope value (e.g., 3\% , corresponding to 10 M$_{\oplus}$). The central contribution of heavy elements is then indeed $m_c$ (provided the parameter $a$ isn't a great deal less than unity; it cannot be more than $\sqrt{\pi}/2$ and presumably everyone would be satisfied with that as the "core mass". In such a case (which is merely chosen to aid the explanation) there would be no sharply defined value of a below which one would say that there is no core. 
\par

Jupiter's primordial internal structure can differ from its present one. Several processes can affect the internal structure during the long-term evolution (timescale of  $\sim$10$^9$ years). The miscibility of heavy elements in metallic hydrogen allows core erosion \citep{Wilson2012}, the extent of which is determined by the ability of overlying thermal convection to do the gravitational work needed to erode the core and develop the gradual redistribution as the planet cools and contracts. Vigorous mixing is expected to occur mostly in the first $10^8$ yrs because that is when the planet cools efficiently and its interior is less degenerate \citep{Guillot2004,Vazan2016}. It seems that in the inner regions where the compositional gradient is steep, the mixing may be inefficient \citep{Vazan2016}.  The efficiency of double-diffusive convection is imperfectly understood and needs further investigation \citep[e.g.,][]{Leconte2012}. 
% then (i.e alpha*T is larger, where alpha is the coefficient of thermal expansion)}

\subsection{Heavy-element accretion during runaway gas accretion}
In both models we present, $M_Z$ at crossover is smaller than the standard estimates of 20-40 M$_{\oplus}$ of heavy element mass for Jupiter from interior models \citep[e.g.,][]{Guillot2005}. Even for $\sigma$=10 g cm$^{-2}$, which is $\sim 3\times$MMSN, $M_Z\sim$16 M$_{\oplus}$. This is even more prominent for $\sigma$=6 g cm$^{-2}$ with M$_Z$ = 7.5 M$_{\oplus}$. Clearly, in order to further enrich Jupiter with heavy elements, more solids (planetesimals) must be accreted during phase-3. Calculating the solid accretion rate during this phase is not trivial since on one hand the planetary feeding zone increases rapidly, while on the other hand, the growing planet is expected to scatter many of the planetesimals.  The exact solid accretion rate during phase-3 is unknown and depends on several physical processes including the accretion morphology, gap formation, planetesimal dynamics, etc. 
\par

Alternatively, Jupiter's envelope could be relatively depleted in heavy elements and have nearly solar composition at least in respect of the elements that are in  
planetesimals. This leaves unanswered the puzzle of why the observed heavy elements in Jupiter's atmosphere exceed solar by a factor of $\sim$ 3. 
It's an open issue as to whether these were delivered by planetesimals. 
A configuration of a core and a proto-solar composition of the envelope has also been suggested by interior modelers \citep[e.g.,][]{Hubbard2016}. In such a scenario, the accreted gas during runaway is expected to be depleted in heavy elements \citep{Helled2014}. 
In any case, it's clear that the solid accretion rate during phase-3 is crucial for determining the final planetary composition. Once the planet reaches a mass of $\sim$ 20 M$_{\oplus}$ most of the accreted mass is hydrogen and helium gas, and as a result, the composition of the gas and/or the solid accretion rates during that stage are crucial. Therefore, in order to link giant planet bulk composition with  origin, we must have a better understanding of the late-accretion. This should include the predicted composition gas, and the composition (and rate) of accreted solids and their formation location in the solar nebula.      
\par

Another way to further enrich the planet with heavies is migration. If Jupiter had formed farther out and then migrated inward, the feeding zone is never depleted and the total heavy element mass can be $\sim$40 M$_{\oplus}$ \citep{Alibert2005}. 
This is different for {\it in situ} formation models of Jupiter in which the planet depletes its feeding zone. 
%A migrating Jupiter on the other hand, is able to accrete more solid material, as its feeding zone is never depleted. 
Further  heavy-element enrichment can also be a result of accretion of gas that is enriched with volatiles. Indeed, \citet{Guillot2006}, suggested that Jupiter's envelope enrichment with noble gases could be due to late accretion of nebular gas depleted in H-He. 

\section{Conclusions and Discussion}
Giant planets' interiors can be more complex than a simple division of a core+envelope. We show that the final composition and structure depend on the conditions available the planet's formation location, and growth history. 
Therefore, Jupiter could consist of a central region in which the heavy element concentration decreases gradually outward.  
While the models we present don't necessarily represent Jupiter's origin, they clearly demonstrate the complexity in modeling Jupiter's origin and the challenges in linking giant planet formation and internal structure. 

We present the sensitivity of the derived planetary composition and internal structure to the model parameters, and emphasize the importance of understanding the late formation stages (phase 3), which determines the final composition of the planetary envelope. However, the primordial distribution of heavy elements could change during the planetary long-term evolution due to core erosion and/or convective mixing \citep[e.g.,][]{Guillot2004, Wilson2012,Vazan2016}.  
We suggest that the mass and composition of Jupiter's core depend on its exact formation history. 
%Overall, it is interesting to note that Jupiter's core mass is expected to be smaller than typically predicted by standard core accretion models.  
We find that the lower $\sigma$ is, the more likely it's to have composition gradients in the deep interior. % of proto-Jupiter. 
While the layers in which the gradient is moderate can mix by convection, the inner region, where the gradient is steep and the heavy element mass fraction is high, is expected to remain stable. 
This configuration could provide the starting point for developing layered-convection in Jupiter \citep{Leconte2012}. Jupiter formation models should be developed further and be combined with the evolution models to investigate whether the material is mixed during Jupiter's long-term evolution.  
It's also desirable to further investigate accretion of both gas and solids and their mixing during phase-3. 
\par

There is no unique definition for JupiterÕs core. How should the core be defined? Is it by the physical state of the material? Its density? Is it the requirement for a density discontinuity? Is it the stability against convection, or is it the fraction of high-Z material that has to be sufficiently large? Defining it by physical state makes little sense- we don't know whether a heavy element-enriched central region would be solid or liquid (it depends on the temperature, which is unknown, and possibly much higher than the adiabat that is typically assumed) and in any event such a definition would de-emphasize the thing we most want to know: The extent to which there is an excess of heavy elements near the planet's center. 
For the same reason, emphasizing the existence or absence of a density discontinuity makes no sense and is not physically meaningful given the quantum mechanical calculations that predict miscibility. This is the essential difference between giant and terrestrial planets, where the sharpness of the core-mantle boundary is a direct consequence of the immiscibility of the core and mantle regions. 
\par

\bibliographystyle{apj}

\clearpage
\renewcommand{\baselinestretch}{1.2}

\begin{figure}
\begin{center}
\includegraphics[angle=0,height=6.8cm]{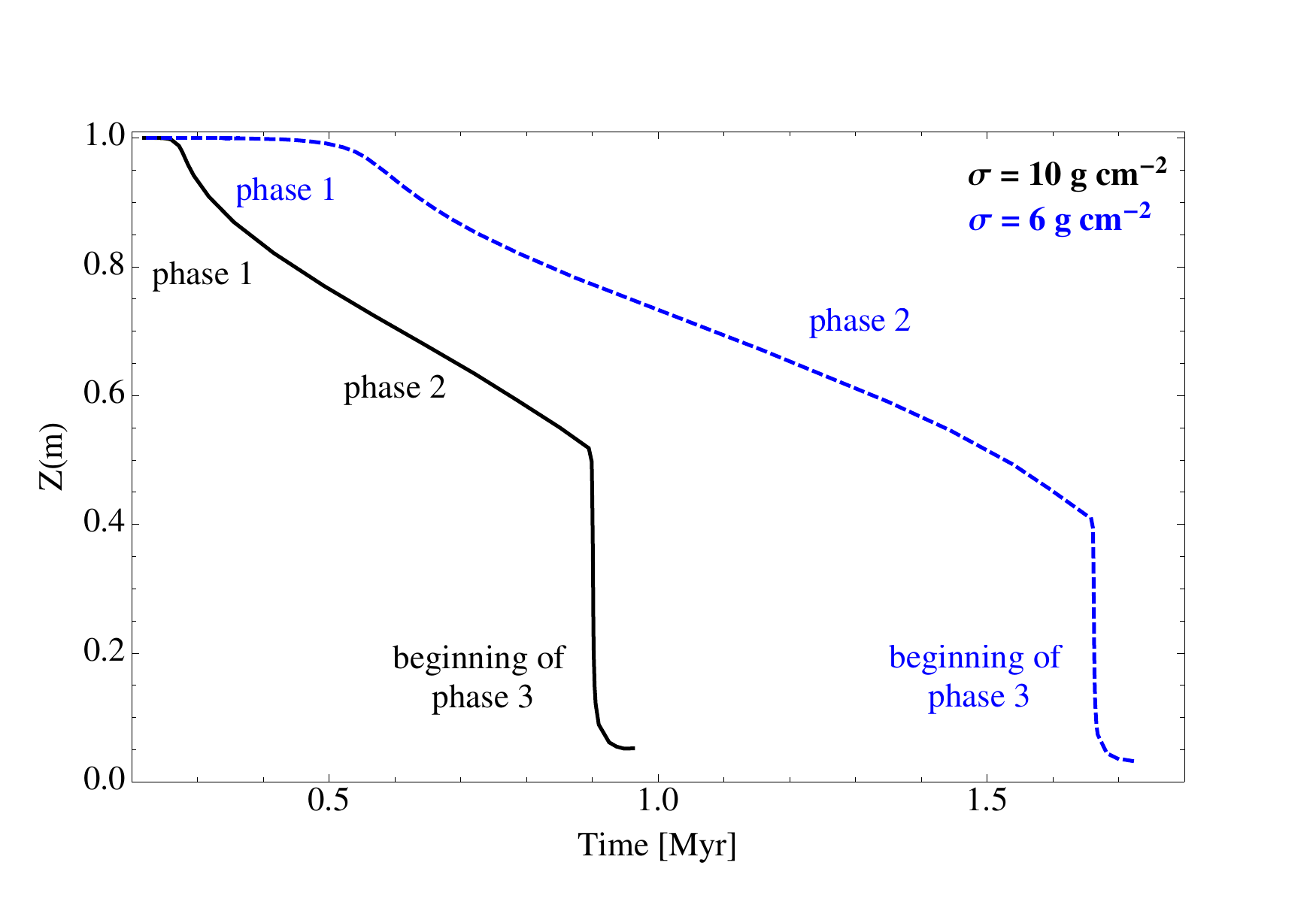}
%\vspace{-1.2cm}
\caption{Z(m) versus time for the two cases $\sigma$ = 10 g cm$^{-2}$ (black) and $\sigma$ = 6 g cm$^{-2}$ (dashed-blue).
%The solid and dashed blue curves correspond to Model-1 and Model-2, respectively. %A zoom-in of Z vs.~time up to a mass of 20 M$_{\oplus}$ is shown in the small panel. 
%This is essentially reflects our claim of Equation 1 on the dependence of the composition on the relative accretion rate, which in our simple case depends solely on the assumed $\sigma$.  
} \label{MMWGFigure}
\end{center}
\end{figure}

\renewcommand{\baselinestretch}{1.2}
\begin{figure}
\begin{center}
\includegraphics[angle=0,height=12cm]{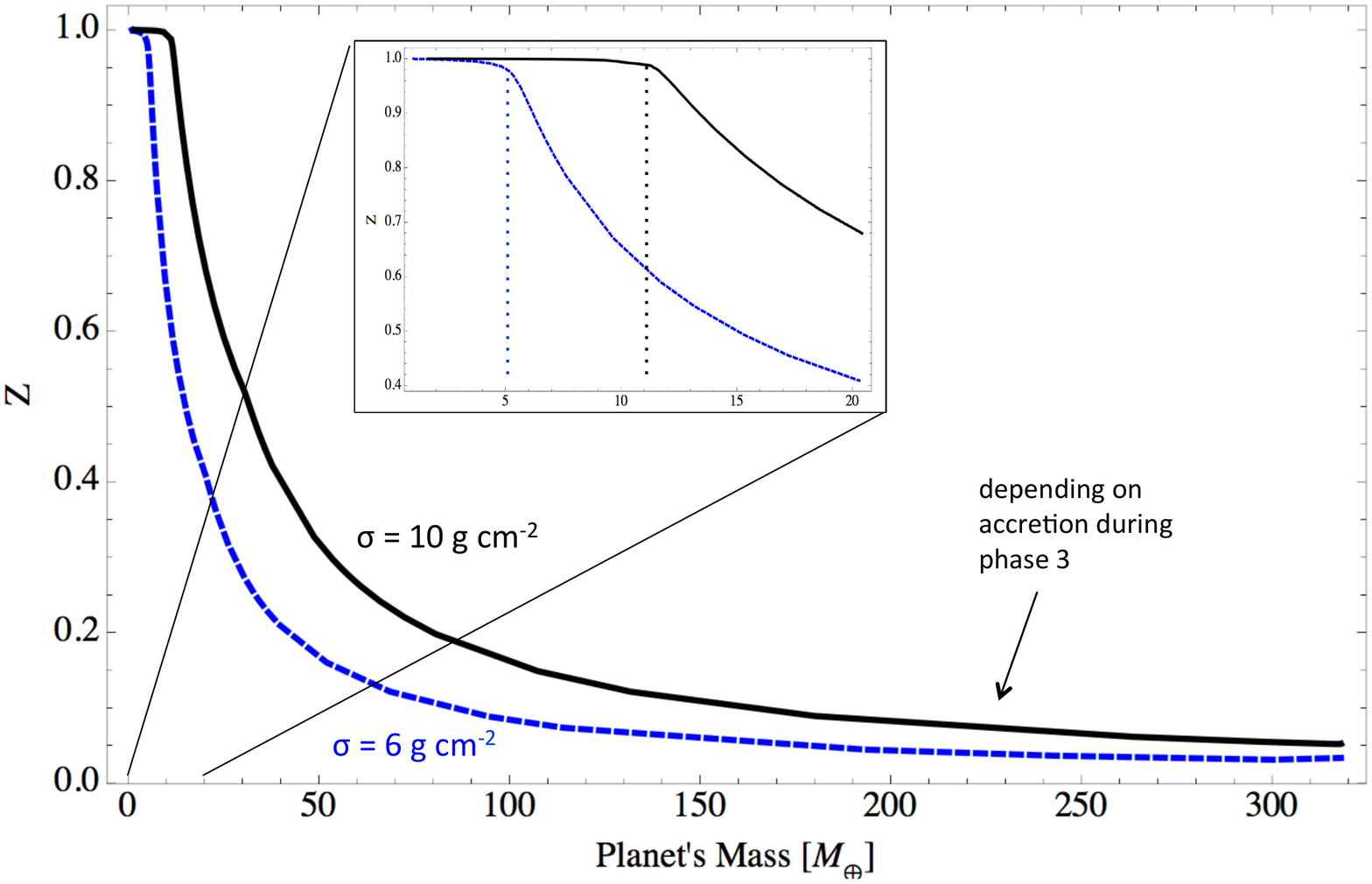}
\vspace{-1.2cm}
\caption{$Z$ versus planetary mass for $\sigma$ = 6 g cm$^{-2}$ (dotted) and $\sigma$ = 10 g cm$^{-2}$ (solid). This demonstrates the dependence of the planetary composition on the relative accretion rate (see Eq.~1). 
A zoom-in of Z vs.~time up to a mass of 20 M$_{\oplus}$ is shown in the small panel.   
} \label{MMWGFigure}
\end{center}
\end{figure}

\renewcommand{\baselinestretch}{1.1}
\begin{figure}
\begin{center}
\includegraphics[angle=0,height=11cm]{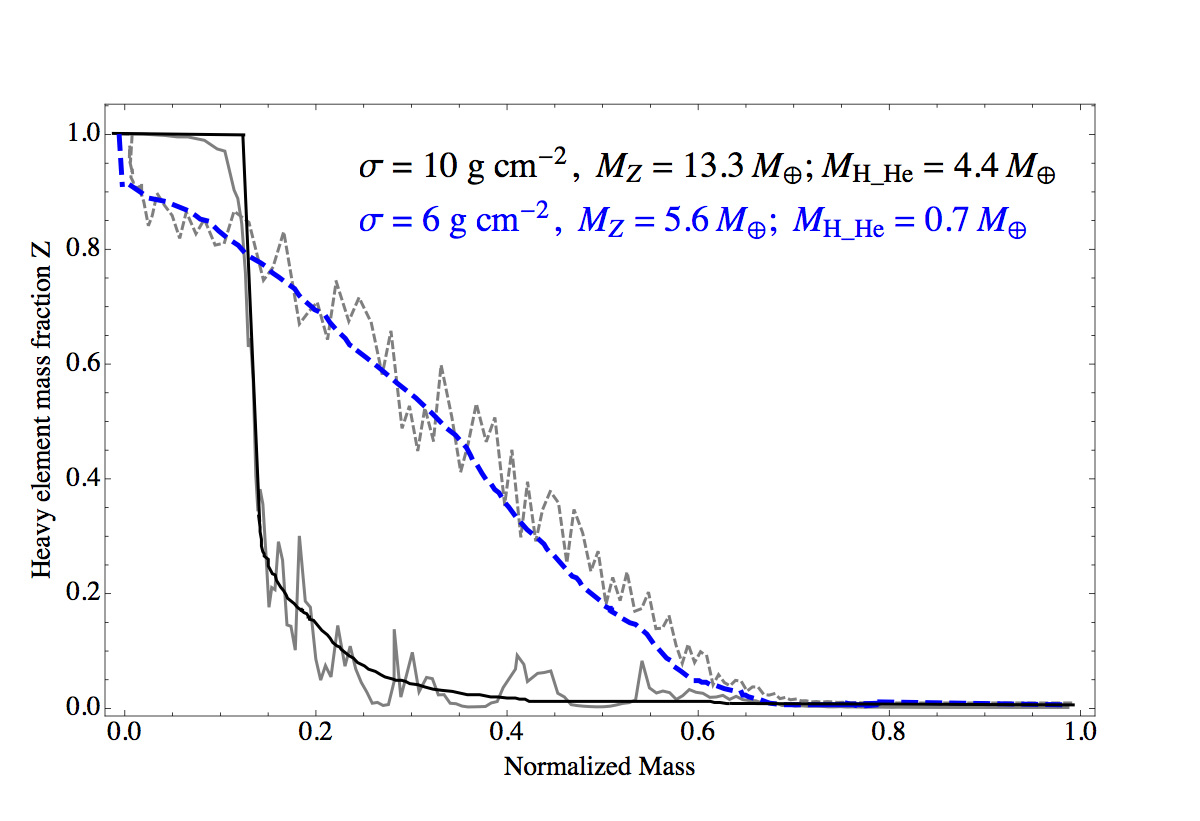}
\caption{$Z$ versus normalized mass for $\sigma$ = 10 g cm$^{-2}$ (solid black) and $\sigma$ = 6 g cm$^{-2}$ (dotted blue) at time of 0.66 Myr.
The two different distributions persist during the planetary formation. Unlike the model of \citep{Lozovsky2017}, we do not assume that mixing and settling take place during the formation process.  
The grey curves show the original distribution before smoothing is applied. The black and blue curves give guidelines to the expected distribution in the two cases. 
}
%It is clear that a lower $\sigma$ results in a more gradual distribution of heavy elements within the planet. } \label{MMWGFigure}
\end{center}
\end{figure}

\end{document}